# Title: Clean 2D superconductivity in a bulk van der Waals superlattice

**Authors:** A. Devarakonda[1], H. Inoue[1†], S. Fang[2‡], C. Ozsoy-Keskinbora[3§], T. Suzuki[1], M. Kriener[4], L. Fu[1], E. Kaxiras[2,3], D.C. Bell[3,5], and J.G. Checkelsky[1*]

**Affiliations:**

[1]Department of Physics, Massachusetts Institute of Technology, 77 Massachusetts Ave., Cambridge, MA 02139, USA.

[2]Department of Physics, Harvard University, 17 Oxford St., Cambridge, MA 02138, USA.

[3]John A. Paulson School of Engineering and Applied Sciences, Harvard University, 29 Oxford St., Cambridge, MA 02138, USA.

[4]RIKEN Center for Emergent Matter Science (CEMS), Wako 351-0198, Japan.

[5]Center for Nanoscale Systems, Harvard University, 11 Oxford St., Cambridge, MA 02138, USA.

†Current address: Institute for Materials Research, Tohoku University, 2-1-1 Katahira, Sendai, Miyagi 980-8577, Japan.

‡Current address: Department of Physics and Astronomy, Center for Materials Theory, Rutgers University, 136 Frelinghuysen Rd., Piscataway, NJ 08854, USA

§Current address: Thermo Fisher Scientific, Achtseweg Noord 5, 5600KA, Eindhoven, The Netherlands

*Correspondence to: checkelsky@mit.edu

**Abstract**:

Advances in low-dimensional superconductivity are often realized through improvements in material quality. Apart from a small group of organic materials, there is a near absence of clean-limit two-dimensional (2D) superconductors, which presents an impediment to the pursuit of numerous long-standing predictions for exotic superconductivity with fragile pairing symmetries. Here, we report the development of a bulk superlattice consisting of the transition metal dichalcogenide (TMD) superconductor 2$H$-niobium disulfide (2$H$-NbS$_2$) and a commensurate block layer that yields dramatically enhanced two-dimensionality, high electronic quality, and clean-limit inorganic 2D superconductivity. The structure of this material may naturally be extended to generate a distinct family of 2D superconductors, topological insulators, and excitonic systems based on TMDs with improved material properties.

**Main Text:**

Superconducting states with real- or momentum-space nodes in the gap function are acutely sensitive to disorder. Such nodal pairing states are weakened by the momentum-space averaging of the gap function caused by disorder scattering of Cooper pairs across the Fermi surface. Conventionally, this disorder averaging can be avoided when Cooper pairs have well-defined crystal momentum, $k$, within the Pippard coherence length, $\xi_0$; a regime realized when $\xi_0$ is smaller than the electronic mean-free path, $\ell$, i.e. $\xi_0/\ell \ll 1$. 2D superconductors in this so-called clean limit play a central role in proposals ranging from archetypal platforms for finite momentum Cooper pairing (*1,2*) to recent constructions for unconventional superconducting phases which leverage normal state spin-textures (*3,4*). At present, there is a paucity of materials in this regime.

Figure 1A presents a historical survey of superconducting materials sorted according to their dimensionality and cleanliness, the former characterized by $H_{c2}^c / H_{c2}^{ab}$ (the ratio of the upper critical magnetic fields perpendicular and parallel to the 2D layer) and the latter by $\xi_0/\ell$. Early experimental work in granular (*g*-Al) and amorphous (*a*-Bi) films demonstrated 2D superconductivity through precise control of the superconducting layer thickness; these systems, subsequently, were used in groundbreaking studies that observed the vortex-antivortex unbinding transition of Berezinskii-Kosterlitz-Thouless (BKT) predicted for 2D superconductors (*5*). Shortly thereafter, studies of similar films revealed the disorder induced superconductor-insulator transition in two-dimensions (*6*). Further developments in thin-film growth have led to the observation of the BKT transition in crystalline films (*7*) (*c*-Nb) and, more recently, to high quality 2D superconductors formed at epitaxial interfaces in LaAlO$_3$/SrTiO$_3$ (*8*) and in δ-doped Nb-SrTiO$_3$ (*9*).

In parallel, the study of anisotropic bulk superconductors has also yielded substantial insight into the nature of the superconducting state. These materials can also be framed in the context of 2D superconductivity (see Fig. 1A); prototypical examples include cuprates, pnictides, organic charge-transfer salts, graphite intercalation compounds, and transition metal dichalcogenides (TMDs). Considerable effort has been invested pursuing tunable dimensionality in these systems. For example, the search for new cuprates in the Ruddlesden-Popper series and their oxygen-deficient variants (*10*) has led to a diverse set of materials with variable dimensionality and a rich variety of exotic behavior (*11*). Discovery of the structurally related layered perovskite Sr$_2$RuO$_4$ realized a 2D stoichiometric superconductor well within the clean limit, in principle satisfying the stability requirements for fragile superconducting gap functions which would otherwise be disrupted by non-magnetic disorder (*12*). Despite the great deal of interest in such phases, thus far this regime of both very clean and highly 2D has largely been limited to organic systems (*13*).

Significant efforts have been put in to achieving 2D superconductivity in transition metal dichalcogenides (TMDs), which given their intrinsically strong spin-orbit coupling and inversion symmetry breaking are expected to yield exotic forms of superconductivity in the clean limit (*14*).

As shown in Fig. 1A for the case of 2$H$-TaS$_2$, intercalation of organic molecules (*15*) ((Py)$_{0.5}$-TaS$_2$) and incommensurate spacer layers (*16*) ((PbS)$_{1.13}$TaS$_2$) has been employed to drive these materials towards the 2D limit and reduce the inversion symmetric coupling between adjacent layers (see associated references in (*17*)). These have led to important advances including establishing that superconductivity is a property of the 2D planes in (Py)$_{0.5}$-TaS$_2$. More recently, developments in the exfoliation of van der Waals (vdW) layered materials have made atomically thin 2D superconductors more readily accessible (e.g. ML-TaS$_2$) (*18–21*). Moreover, subsequent studies have revealed a form of superconductivity characterized by strong Ising spin-orbit coupling equivalent to applying magnetic fields of order 100 T (*19–21*). However, flakes exfoliated from the bulk are often subject to degradation and reduction in quality during the fabrication process (*22*). Here we show that high-quality $H$-NbS$_2$ monolayers with electronic mobilities more than three orders of magnitude larger than in bulk 2$H$-NbS$_2$ can be realized in a bulk single crystal superlattice formed with a commensurate block layer. Correspondingly, we find that this material is a clean limit 2D superconductor exhibiting a BKT transition at $T_{BKT} = 0.82$ K and prominent 2D Shubnikov de-Haas (SdH) quantum oscillations.

The fundamental structural unit in hexagonal TMDs is the $H$-$MX_2$ layer where $M$ and $X$ are a transition metal and chalcogen, respectively. As shown in Fig. 1B, this structure (point group symmetry $\bar{6}m2$ ($D_{3h}$)) breaks inversion symmetry in the layer plane owing to the trigonal prismatic coordination of $X$ around $M$ (the missing inversion partners are shown as dashed circles), and as a result yields an out-of-plane (Ising) spin-texture (*18–21*). For thin flakes deposited on substrates, the substrate-flake interface breaks mirror symmetry (Fig. 1C) and yields an in-plane (Rashba) spin-texture (*23*). Taken together, for TMD flakes on substrates, the simultaneous breaking of mirror and inversion symmetry leads to a mixed spin texture (Fig. 1D) on the Fermi surface composed of both Ising and Rashba components (*18*). These spin-textures, and the resulting physics are suppressed in the bulk limit where the overall unit cell preserves inversion symmetry.

We have synthesized a single crystal material $Ba_6Nb_{11}S_{28}$ composed of high-quality *H*-$NbS_2$ layers and $Ba_3NbS_5$ block layers in which the TMD layers are strongly decoupled (see section 2 of (*17*) for discussion of material modeling and structure). Figure 1E shows a cross section of the structure imaged by high angle annular dark field scanning transmission electron microscopy (HAADF-STEM) with the model structure superimposed. As determined by electron and powder x-ray diffraction, the unit cell (space group $P\bar{3}1c$ with $a = 10.4$ Å, $c = 24.5$ Å) is composed of two inversion-related *H*-$NbS_2$ layers across each of which mirror symmetry is broken by the neighboring block layers; the overall unit cell retains inversion symmetry. The *H*-$NbS_2$ interlayer distance $d = 8.9$ Å is more than three times that of 2*H*-$NbS_2$ (*24*), leading to a reduction of the interlayer transfer integral $t_\perp$. This amplifies the two-dimensionality of the electronic structure relative to 2*H*-$NbS_2$ and enables local symmetry breaking induced spin-orbit textures on the *H*-$NbS_2$ layers (*25*, *26*) (*H*-$NbS_2$ monolayers in $Ba_6Nb_{11}S_{28}$ experience local inversion symmetry breaking with point group symmetry 32 ($D_3$), see section 5(e) of (*17*)). Compared to traditional misfit compounds, which combine incommensurate layers in a superlattice, $Ba_6Nb_{11}S_{28}$ exhibits a $3 \times 3$ in-plane, commensurate superstructure caused by the difference in in-plane lattice constants between the two layer types (see TEM diffraction linescans in section 2(a) of (*17*)), which leads to additional modification of the electronic structure.

Figure 1F shows the temperature, $T$, dependence of electrical resistivity, $\rho_{xx}(T)$, for $Ba_6Nb_{11}S_{28}$. The system is a metal, eventually showing superconductivity below $T = 1$ K. This can be compared to bulk 2$H$-NbS$_2$ which is also metallic and becomes a superconductor at $T_c = 5.7$ K. Unlike several other related $H$-$MX_2$ systems, neither Ba$_6$Nb$_{11}$S$_{28}$ nor 2$H$-NbS$_2$ show signs of a density wave transition (*27*) (see section 12(e) of (*17*) for associated discussion). The inset of Fig. 1F shows a detailed view of the superconducting transition, which onsets near $T = 1.6$ K and reaches zero resistance at $T = 0.85$ K. At the latter temperature, the magnetic susceptibility $4\pi\chi$ with field along the $c$-axis shows a Meissner signal reaching a shielding fraction of 75% (Fig. 1F inset, dark green ZFC curve) and a volume fraction of 40% (Fig. 1F inset, light green FC curve). The reduction in the transition temperature for the NbS$_2$ layers compared to the bulk is similar to that observed in organic intercalated variants (*28*) and is consistent with a reduction of the electron-phonon coupling strength inferred from the measured resistivity at high $T$ (see section 12(d) of (*17*)).

The temperature dependence of resistivity for $0.85$ K $< T < 1.6$ K is well described by the Halperin-Nelson model, $\rho_{xx}^F(T) = \rho_{xx}^N e^{-b/\sqrt{t}}$, where $\rho_{xx}^F$ and $\rho_{xx}^N$ are the fluctuation and normal state resistivity respectively, $t = T/T_{HN} - 1$, and $b$ is a fitting parameter on the order of one (dashed curve in Fig. 1F inset) (*29*). The agreement with the Halperin-Nelson model evidences fluctuations of the superconducting order parameter above a two-dimensional BKT transition. Such behavior is generally rare in bulk single crystals, but has been reported in La$_{1.875}$Ba$_{0.125}$CuO$_4$ and attributed to the decoupling of superconducting CuO$_2$ planes by stripe order (*30*). This is a further departure from the behavior in bulk 2$H$-NbS$_2$, which exhibits a sharp superconducting transition (see section 12(a) of (*17*) and fig. S23); instead it closely resembles those observed in monolayer $H$-$MX_2$. Further evidence of increased two-dimensionality is the low $T$ resistivity anisotropy $\rho_{zz}/\rho_{xx} > 10^3$ in the normal state of Ba$_6$Nb$_{11}$S$_{28}$ ($\rho_{zz}$ is the $c$-axis resistivity), which is substantially enhanced compared to $\rho_{zz}/\rho_{xx} \sim 100$ in 2H-NbS$_2$ (see section 12(b) of (*17*) and fig. S24).

Magnetotransport measurements demonstrate the cleanliness of this material and show further evidence for a 2D electronic structure. Figure 2A shows the magnetoresistance $MR \equiv (\rho_{xx}(H)/\rho_{xx}(0)) - 1$, measured to 31 T. We observe SdH quantum oscillations that respond to the component of the magnetic field perpendicular to the *ab*-plane (the tilt angle $\theta$ is measured between the *c*-axis and applied field). The Fast Fourier Transform (FFT) computed after subtracting a monotonically increasing background (see section 5(b) of (*17*) and fig. S14) plotted versus inverse field shows this more clearly (Figs. 2, B and C). Here the oscillation frequency multiplied by $\cos(\theta)$ has little variance versus angle demonstrating the 2D nature of the Fermi surface. This is qualitatively different than in 2H-NbS$_2$, for which electronic structure calculations indicate warped and elliptical Fermi surfaces (*31*). Owing to the reduced coupling between the TMD layers, the observed bands (labeled here as $\alpha$, $\beta_{(1,2)}$, and $\gamma_{(1,2)}$) can be understood by starting with the 2D electronic structure of monolayer H-NbS$_2$, which consists of bands at the $\Gamma$, *K*, and *K'* points of the hexagonal Brillouin zone (BZ) (Fig. 2D), and zone-folding into a reduced BZ determined by the $3 \times 3$ superstructure imposed by the block layers (see Fig. 2E). In particular, the approximate order of magnitude reduction in the pocket size from monolayer H-NbS$_2$ caused by this zone-folding quantitatively captures the size of the observed pockets, Fig. 2F, and is further supported by first-principles calculations (see section 4 and 5(c) of (*17*) for additional discussion and fig. S12). An important aspect of this structure is that the large ratio of the spin-orbit coupling to $t_\perp$, evident from the degree of two-dimensionality, enables local symmetry breaking to affect the bulk electronic structure. The zone-folding promotes the Rashba-textured pockets associated with the Γ point in monolayer $H$-NbS$_2$ to be larger than the Ising-split pockets at $K$ and $K'$ (supported by comparing the calculated band structure for monolayer $H$-NbS$_2$ with that of the $3\times3$ zone-folded structure, section 4 and 5(e) of (*17*), and figs. S11 and S12) and has potential implications for superconducting pairing.

More generally, it is noteworthy that quantum oscillations have not been reported in 2$H$-NbS$_2$; there, the typical transport mobilities reported for bulk single crystals are of order 1 cm$^2$ / V s (*27*). In Ba$_6$Nb$_{11}$S$_{28}$ we see the onset of SdH quantum oscillations in magnetic fields between $2-3$ T, indicating quantum mobilities of order $10^3$ cm$^2$ / V s. Analysis of the quantum oscillations and low field magnetoresistance indicates an associated transport mean free path $\ell = 1.21$ μm greatly exceeding the Pippard coherence length $\xi_0 \approx 0.18\hbar v_F / k_B T_c = 254$ nm (see section 7 of (*17*)), placing Ba$_6$Nb$_{11}$S$_{28}$ in the clean limit of superconductivity (Fig. 1A).

Turning to properties of the superconducting state, Fig. 3A shows the current voltage $I(V)$ characteristics of Ba$_6$Nb$_{11}$S$_{28}$ across the superconducting transition. As expected for a BKT transition (*29*), a linear response at $T = 0.95$ K and above crosses over to a non-linear dependence $V \propto I^\alpha$ with $\alpha \sim 3$ at $T_{BKT} = 0.82$ K, consistent with $T_{HN} = 0.85$ K. With further examination of the fluctuation conductivity and the slope of the power law exponent close to $T_{BKT}$, we find evidence for a vanishingly small interlayer coupling in the superconducting state (*32*) (see section 6 of (*17*)). Figure 3B shows the evolution of $\rho_{xx}(H)$ as a function of magnetic field for different θ. Whereas for $\theta = 0°$ superconductivity is suppressed with relatively low fields and gives rise to quantum oscillations, for larger θ the upper critical field $\mu_0 H_{c2}$ rapidly increases (herein we define $\mu_0 H_{c2}$ and its error to be when $\rho_{xx}$ reaches 50 ± 5 percent of the normal state value).

Figure 3C summarizes this behavior with $\mu_0 H_{c2}(\theta)$ showing a sharp cusp for in-plane fields. Recent studies of 2D superconductors have demonstrated that a distinguishing feature of such systems from anisotropic 3D superconductors is the profile of $\mu_0 H_{c2}(\theta)$ following the 2D Tinkham form $\left(\frac{H_{c2}(\theta)\sin\theta}{H_{c2}^{ab}}\right)^2 + \left|\frac{H_{c2}(\theta)\cos\theta}{H_{c2}^{c}}\right| = 1$, where $H_{c2}^{ab}$ and $H_{c2}^{c}$ are the upper-critical fields for field applied in-plane and out-of-plane respectively (33). The response of Ba$_6$Nb$_{11}$S$_{28}$ can be fit by such a form, contrasting the anisotropic 3D character of 2H-NbS$_2$ (see section 12(c) of (17) and fig. S25). Furthermore, for Ba$_6$Nb$_{11}$S$_{28}$, we observe an enhancement of the scale of $\mu_0 H_{c2}(\theta)$ for angles below 1.7° measured relative to the *ab*-plane. As shown in the inset of Fig. 3C, this anomalous enhancement coincides with $\mu_0 H_{c2}(\theta)$ crossing the Pauli paramagnetic limit $\mu_0 H_P \approx 1.84\, T_{BKT} = 1.51$ T. We find that two independent Tinkham fits trace the data across the entire angular regime: for $|\theta - 90°| \geq 1.7°$ we have $\mu_0 H_{c2}^{c} = 0.15$ T, $\mu_0 H_{c2}^{ab} = 2.19$ T and for $|\theta - 90°| \leq 1.7°$ we obtain $\mu_0 H_{c2}^{c} = 0.09$ T, $\mu_0 H_{c2}^{ab} = 2.55$ T.

To further examine the anomaly in $\mu_0 H_{c2}$, we measured $\rho_{xx}(H,T)$ with $\theta$ systematically tuned away from 90°. Plotted as the excess conductivity $\delta\sigma \equiv 1 - \rho_{xx}/\rho_{xx}^{N}$, the substantial enhancement at low $T$ and high $H$ quickly disappears as $\theta$ is moved away from 90°, and by $\theta = 86°$ there is little variation with further field tilt (Fig. 4A). A distinct feature at all $\theta$ is the finite $\delta\sigma$ associated with fluctuating superconductivity for low $H$ extending to $T$ beyond $T_{BKT}$. To remove this fluctuation contribution, we plot the difference $\delta\sigma(\theta = 90°) - \delta\sigma(\theta = 84°)$ in Fig. 4B. The expected 2D Ginzburg-Landau (2D-GL) behavior is shown as a green line; the transition line follows this response below $T_{BKT}$ until approximately $T/T_{BKT} \approx 0.6$, below which a considerable enhancement is observed. As shown in Fig. 4C, this behavior is confined to low temperature and to a small angular region $\delta\theta$ about the *ab*-plane.

These observations taken together indicate the appearance of a clean 2D superconducting state with enhanced stability (larger $H_{c2}$) when $T < 0.5\, T_c$ and field $H > H_P$ applied very close to the layer plane. Various theoretical scenarios have been discussed for Pauli breaking in 2D superconductors including spin-orbit scattering (*34*), Ising superconductivity (*19–21*), and Fulde-Ferrell-Larkin-Ovchinnikov (FFLO) states (*1, 2*). Given the clean limit superconductivity realized here, spin-orbit scattering enhancements cannot account for the present observations. The dominant local Rashba spin-orbit coupling in the present system reduces the importance of the local Ising coupling (*35*) and in Ising superconductors no abrupt change in $H_{c2}(\theta)$ is expected. Instead, the acute angle-dependent enhancement resembles that of the clean layered organic FFLO candidate $\beta''$-$(ET)_2SF_5CH_2CF_2SO_3$ in similar conditions of $H$ and $T$ arising when the orbital-limiting of superconductivity is quenched by aligning $H$ close to the layer plane (*36*). As shown schematically in the inset to Fig. 4C, once orbital-limiting is overcome within a critical angular window $\delta\theta$ (typically on the order of $1-2°$ in organic systems (*37*)), one naturally expects a sharp enhancement of $H_{c2}(\theta)$ as the large in-plane magnetic field stabilizes a finite-momentum pairing state.

The close parallel between our observations and those expected for a 2D FFLO phase (along with our ability to fit the phase diagram in Fig. 4B with predictions for such a phase, section 10 of (*17*)) highlight it as a promising candidate. Theoretical studies of 2D FFLO superconductors further predict a cascade of magnetic vortex states with finite-momentum for $T/T_c < 0.55$, $H > H_P$, and $|\theta - 90°| < \delta\theta$ (see Fig. 4C (inset, red dashed-line)), which may be resolved by higher resolution measurements. More directly, compared to clean 2D organic superconductors (Fig. 1A), the robust inorganic nature of $Ba_6Nb_{11}S_{28}$ offers the opportunity to examine the potential real space modulation of superconductivity using, for example, scattering techniques (*38*). Compared to other clean inorganic systems, $Ba_6Nb_{11}S_{28}$ is derived from a well-studied family of materials, lacks localized magnetic moments, and perhaps most importantly, has vdW layer bonding that allows for exfoliation and integration into superconducting device structures. One exciting consequence could be simplified fabrication of phase-sensitive junction devices similar to those used to study the cuprates and other unconventional superconductors (*39*).

We hypothesize that the large enhancement of electronic mobility observed for the *H*-$NbS_2$ layers in $Ba_6Nb_{11}S_{28}$ may be attributed to screening by the highly polarizable block layer akin to that observed in engineered semiconductor heterostructures (*40*). Additionally, our density-functional theory (DFT) calculations suggest that the lowest energy cleavage occurs between the *H-MX₂* and block layers implying that mechanically exfoliated $Ba_6Nb_{11}S_{28}$ may yield naturally encapsulated *H*-$NbS_2$ monolayers akin to vdW structures made by stacking *MX₂* layers and h-BN (*22*) (we observe using optical and atomic-force microscopy that standard exfoliation techniques can be used to obtain flakes suitable for device fabrication, section 11 of (*17*)). However, given that bulk $Ba_6Nb_{11}S_{28}$ already exhibits two-dimensional physics, we propose that insertion of commensurate spacer layers could be an alternative to fabricating exfoliated nanodevices. There is also scope for functionalizing the spacer layer to further modulate the *H-MX₂* layer, for example by introducing magnetic constituents. The large electronic mean-free path of $Ba_6Nb_{11}S_{28}$ enables clean-limit superconductivity and can potentially realize unconventional phases predicted in monolayer *H-MX₂* superconductors (*3, 4, 41, 42*). The physics of other *MX₂* materials that has captured considerable attention can also benefit from longer electron mean-free paths. In particular, extending the materials family of $MX_2$ natural, commensurate superlattices may, for example, pave the way to longer mean-free paths enabling topological edge mode circuitry in $WTe_2$ superlattices (*43*) or longer exciton lifetimes in $MoS_2$ and related semiconducting TMD materials (*44*).

**Acknowledgments:** We are grateful to Y. Tokura, P. A. Lee, M. Nakano, H. Matsuoka, N.F.Q. Yuan, V. Mitrovic, and S.-L. Zheng for fruitful discussions and to M. Kamitani and A. Akey for technical support. **Funding:** This research is funded in part by the Gordon and Betty Moore Foundation through grants GBMF3848 and GBMF9070 to JGC (instrumentation development), the Office of Naval Research (ONR) under Award N00014-17-1-2883 (advanced characterization), and the U.S. Department of Energy (DOE), Office of Science, Basic Energy Sciences (BES), under Award DE-SC0019300 (material development). A.D., S.F., C.O.-K., and E. K. acknowledge support by the STC Center for Integrated Quantum Materials, NSF Grant No. DMR-1231319. L.F. acknowledges support by the DOE Office of Basic Energy Sciences under Award No. DE-SC0018945. Computations were performed on the Odyssey cluster supported by the FAS Division of Science, Research Computing Group at Harvard University, and the Extreme Science and Engineering Discovery Environment (XSEDE), which is supported by National Science Foundation grant number ACI-1548562. A portion of this work was performed at the National High Magnetic Field Laboratory, which is supported by National Science Foundation Cooperative Agreement No. DMR-1157490, the State of Florida, and the US Department of Energy. **Author contributions:** A.D. synthesized and characterized the single crystals. A.D. and H.I. performed the electrical transport experiments. A.D. and M.K. performed the magnetization experiments. C.O.-K. and D.B. performed the electronic microscopy experiments. A.D. and S.F. performed theoretical calculations. All authors contributed to discussions and writing the manuscript. J.G.C. coordinated the project. **Competing interests:** The authors declare no competing interests. **Data and materials availability**: The data in the manuscript are available from the Harvard Dataverse (45).

**Figure 1: 2D superconductivity and $Ba_6Nb_{11}S_{28}$**

(**A**) Survey of superconducting materials characterized by anisotropy of the upper critical field $H_{c2}^{c}/H_{c2}^{ab}$ and ratio of the Pippard coherence length to mean free path $\xi_0/\ell$. The boundary between the clean and dirty limits is shown as a horizontal line. (**B**) Crystal structure of $H$-$MX_2$ projected onto the $ab$-plane. Lack of inversion-symmetry is illustrated by the missing chalcogen ($X$) inversion partners (dashed circles). (**C**) The $ab$-plane mirror symmetry in monolayer $H$-$MX_2$ can be broken by substrates or local fields ($\nabla U$). (**D**) Depiction of momentum space spin-orbit texture for monolayer $H$-$MX_2$ with varying degrees of Ising and Rashba coupling. (**E**) HAADF-STEM image of $Ba_6Nb_{11}S_{28}$ taken along $[1\bar{1}00]$ axis (1 nm scale bar). A simulation of the model structure is overlaid with one unit cell shaded in green. (**F**) Resistivity as a function of temperature $\rho_{xx}(T)$ in $Ba_6Nb_{11}S_{28}$ showing the superconducting transition. Upper inset: Magnified view of the transition in $\rho_{xx}(T)$ and magnetic susceptibility $4\pi\chi$ measured with zero-field cooling (ZFC) and field-cooling (FC). $\rho_{xx}(T)$ is well-fit by the Halperin-Nelson model shown in black (see text). Lower inset: $H$-$NbS_2$ layer and mirror symmetry breaking $Ba_3NbS_5$ block layers.

**Figure 2: Quantum oscillations and electronic structure of $Ba_6Nb_{11}S_{28}$**

(**A**) Magnetoresistance as a function of perpendicular field $MR \equiv (\rho_{xx}(\mu_0 H_\perp)/\rho_{xx}(0))-1$ at temperature $T = 0.39$ K for different field rotation angles $\theta$ (geometry defined as shown in the inset). Curves are vertically offset by 150% of MR for clarity. (**B**) Low frequency and (**C**) full range of quantum oscillation amplitude Fast Fourier Transform (FFT) as a function of perpendicular frequency $F\cos(\theta)$. The FFT amplitude for the higher frequency pockets are multiplied by 25. (**D**) DFT calculation of monolayer $H$-$NbS_2$ Fermi surfaces including spin-orbit coupling (*17*). **(E)** Depiction of zone-folding scheme involving the $3\times3$ superstructure imposed by the Ba$_3$NbS$_5$ block layer where the reduced Brillouin zone is enclosed by the bold line. **(F)** Electronic structure of zone-folded monolayer *H*-NbS$_2$ with experimentally observed Fermi surface cross-sectional areas drawn to scale as filled circles. The black box corresponds to 0.01 Å$^{-2}$.

**Figure 3: 2D superconductivity and Pauli limit breaking in Ba$_6$Nb$_{11}$S$_{28}$**

**(A)** Current-voltage characteristics $I(V)$ from temperature $T = 0.95$ K to $T = 0.28$ K. The inset shows the evolution of the power law $V \propto I^\alpha$ with the horizontal line marking $\alpha = 3$. **(B)** Longitudinal resistivity $\rho_{xx}$ as a function of field $\mu_0 H$ for different $\theta$. Curves are vertically offset by 20 µΩ cm for clarity (horizontal lines). Vertical ticks separate regions measured with low current ( 7 µA ) and higher current ( 70 µA ) to avoid Joule heating suppression of superconductivity. For $\theta = 80°$ and $90°$, only low current is used. **(C)** Angular dependence of upper critical field $\mu_0 H_{c2}$ measured at $T = 0.28$ K with fits to the 2D-Tinkham model, purple and black lines, computed using data in the range $|\theta-90°|<1.7°$ and $|\theta-90°|>1.7°$, respectively. The inset shows a detailed view near $\theta = 90°$ where an enhancement of $\mu_0 H_{c2}(\theta)$ is observed across the Pauli limit $\mu_0 H_P$.

**Figure 4: Superconducting phase diagram of Ba$_6$Nb$_{11}$S$_{28}$**

**(A)** Excess conductivity relative to the normal state $\delta\sigma(\mu_0 H, T)$ for field angles $\theta$ near the *ab*-plane ($\theta = 90°$). **(B)** Difference between $\delta\sigma(\mu_0 H, T)$ for $\theta = 90°$ and $84°$. The temperature axis is normalized to $T_{BKT}$. The 2D Ginzburg-Landau model of $\mu_0 H_{c2}$ is shown in green. **(C)** The angular dependence of $\mu_0 H_{c2}$ at $T/T_{BKT} = 0.3$ (orange) and $\mu_0 H_{c2}$ at $T/T_{BKT} = 0.8$ (green, magnified by a factor of 3). Inset: Schematic depiction of $\mu_0 H_{c2}$ in a clean 2D system where an enhancement is expected within a critical region $|\theta - 90°| < \delta\theta$ where orbital-limiting is quenched and an FFLO state appears for $T/T_c < 0.55$ and $H > H_P$ (dark blue, solid line). Theoretical studies of 2D FFLO superconductors further predict a cascade of magnetic vortex states that appear as a corrugation of $\mu_0 H_{c2}(\theta)$ within this regime (*37*) (red, dashed line).

**Supplementary Materials:**

Materials and Methods

Supplementary Text

Figures S1-S26

Tables S1-S6

References (46-154)

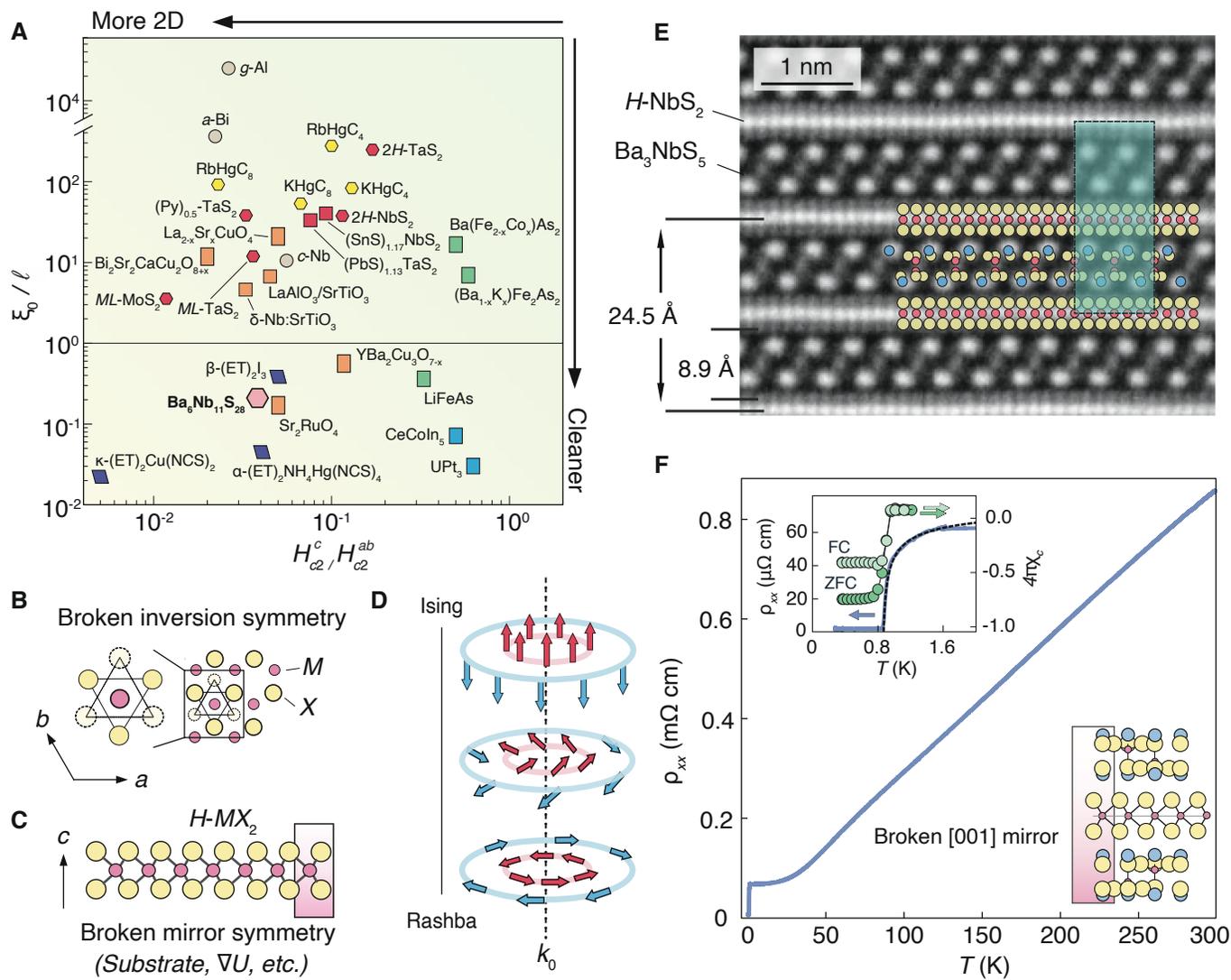

Fig. 1. Devarakonda et al.

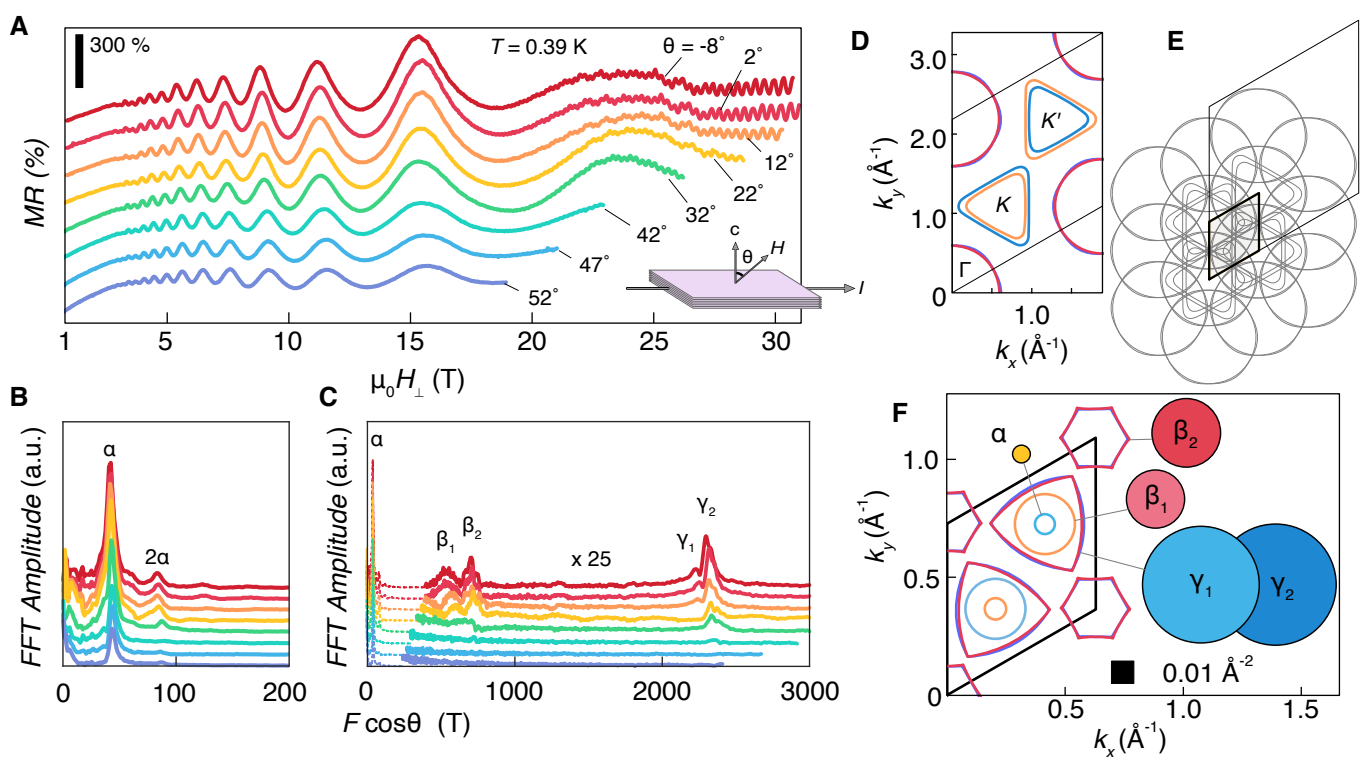

Fig. 2. Devarakonda et al.

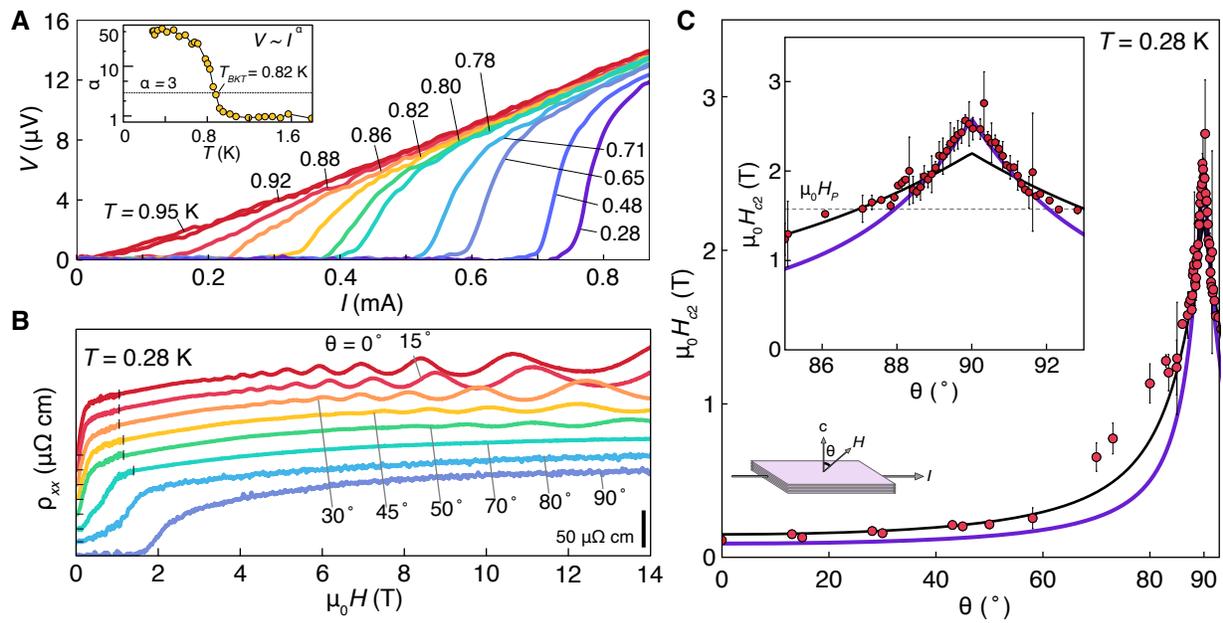

Fig. 3. Devarakonda et al.

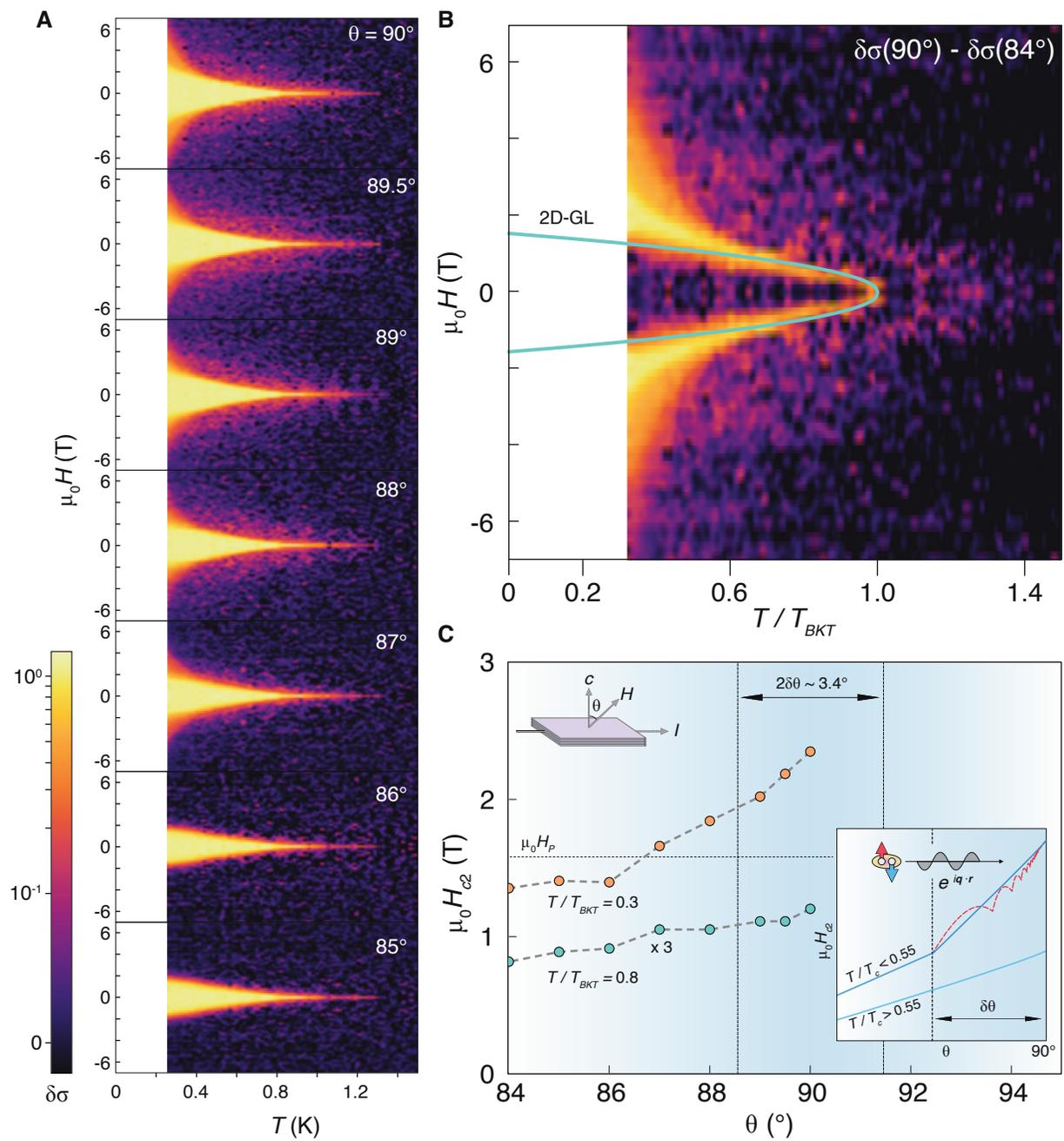

Fig. 4. Devarakonda et al.